\documentclass[journal]{IEEEtran}
\usepackage{amsmath,amssymb,amsthm}
\usepackage{graphicx}
\usepackage{tikz}
\usetikzlibrary{positioning,arrows.meta}
\usepackage{booktabs}
\usepackage{cite}
\usepackage[hidelinks]{hyperref}

\newtheorem{theorem}{Theorem}
\newtheorem{proposition}{Proposition}
\newtheorem{corollary}{Corollary}
\theoremstyle{remark}
\newtheorem{remark}{Remark}

\newcommand{\eul}[2]{\genfrac{\langle}{\rangle}{0pt}{}{#1}{#2}}
\newcommand{\stir}[2]{\genfrac{\{}{\}}{0pt}{}{#1}{#2}}

\newcommand{\AuthorAffiliation}{SKAIChips Co., Ltd.}

\title{Time-Domain Rate-Invariant Comb Decimators\\ by Cascading the Euler--Frobenius Polynomial}
\author{Ealwan~Lee%
\thanks{E. Lee is with \AuthorAffiliation}}
\begin{document}
\maketitle

\begin{abstract}
The pass-band droop of an $N$th-order comb (CIC) decimator depends on the
decimation factor $M$, and so do the closed-form coefficients of its low-rate
droop compensator. A recent method removes this dependency to first order
with a symmetric three-tap filter in the integrator section. This paper shows
that the three-tap filter is a low-order approximation of an exact one: the FIR filter
whose coefficients are the Eulerian numbers of order $N-1$, that is, the
Euler--Frobenius polynomial. With it, the impulse response of the comb
equals, sample for sample, the B-spline of degree $N-1$ sampled at $M$ points
per knot, for every $M$. The filter is also the unique one for which doubling
the rate leaves the response unchanged at the common sampling instants; this
time-domain rate invariance is exact. Three consequences follow. The
closed-form coefficients of maximally flat compensators become functions of
$N$ alone. The group delay is exactly $N/2$ low-rate samples for every $M$,
for even and odd $N$ alike. The residual rate dependency of the pass band is
given in closed form and decays as $M^{-N}$ or $M^{-(N+1)}$, against $M^{-4}$
for the three-tap filter and $M^{-2}$ for the plain comb. The underlying
identity is equivalent to the discrete B-spline factorization of spline
signal processing; the contribution is its identification with Eulerian
numbers and its application to comb filters. The aim is rate invariance, not
a flatter pass band: for $N\le4$ the residual rate dependency is no smaller
than with the three-tap design, and what the exact filter adds is exactness in
the time domain.
\end{abstract}


\section{Introduction}

Comb decimators \cite{chu}, implemented as cascaded integrator--comb (CIC)
structures \cite{hogenauer}, are the standard first stage of high-ratio decimation chains
because they need no multipliers and their structure does not change with the
decimation factor $M$. Their pass-band droop is usually corrected by a short
compensation filter at the low rate
\cite{kim,dolecek,molnar}. The droop, however, depends on $M$ as well as on the
order $N$, and so do the closed-form coefficients of the compensator: a system
that supports several rates must store or recompute them for each $M$.

In \cite{lee} it was observed that the $M$-dependent part of the droop can be
cancelled, to first order in $M^{-2}$, by a symmetric three-tap filter placed
in the integrator section, with coefficients that depend on $N$ alone. After
this \emph{derating} filter the compensator sees a droop that is nearly
independent of $M$. The word is meant in both of its senses. For finite $M$ the comb droops less
than its limit for $M\to\infty$; the filter \emph{derates} it in the
engineering sense, lowering the response for every $M$ to that worst case.
Having done so, it \emph{de-rates} it: the droop no longer depends on the
rate. As in \cite{lee}, the term is used exclusively for the pass-band
deviation against a change of $M$; \emph{compensation} is reserved for the
correction of the droop itself, which derating does not attempt.

This paper answers the question that \cite{lee} leaves open: what is the filter
that removes the rate dependency exactly, rather than to first order? The
answer is the filter whose coefficients are the Eulerian numbers
\cite{gkp,petersen}. The result follows from a classical identity of Euler,
and it is equivalent to the factorization of discrete B-splines derived by
Unser, Aldroubi and Eden for image interpolation \cite{unser91,unser93a}. That
work, however, does not consider comb decimators, rate dependency or
compensator design, and does not identify the coefficients as Eulerian
numbers; conversely, the CIC literature does not appear to use the identity.
The purpose of this paper is to connect the two.

The contributions are as follows.
\begin{itemize}
\item The exact derating filter $E_N(z)$ is given in closed form for every
order (Theorem~\ref{thm:main}), with a self-contained proof from Euler's
identity.
\item $E_N$ is shown to be the \emph{unique} correction filter for which
doubling the rate leaves the response unchanged at the common instants
(Theorem~\ref{thm:ri}); the filter follows from this time-domain
rate-invariance requirement alone, and the invariance is exact.
\item The residual rate dependency of the pass band is derived in closed form,
together with its leading term for the comb, the three-tap filter of
\cite{lee} and the exact filter (Proposition~\ref{prop:eps}).
\item It is shown that the closed-form coefficients of maximally flat
low-rate compensators lose their dependence on $M$ (Section~\ref{sec:comp}), and that the group delay is the same
number of low-rate samples for every $M$ and every $N$
(Corollary~\ref{cor:prop}).
\item The three-tap filter of \cite{lee} is shown to coincide with a member of
the exact family for $N=3$ and $N=4$, and to be its second-moment
approximation otherwise.
\item A multiplier-free realization that taps the integrator chain, with
weights given by Stirling numbers, is described.
\end{itemize}

\section{Background}

\subsection{Comb decimator and its rate dependency}

The $N$th-order comb filter with decimation factor $M$ is
\begin{equation}
H_{N,M}(z)=\left(\frac{1-z^{-M}}{1-z^{-1}}\right)^{N},
\label{eq:comb}
\end{equation}
operating at the high (input) rate and followed by down-sampling by $M$.
Throughout, $\omega$
denotes angular frequency normalized to the \emph{low} rate and
$\theta=\omega/M$ the same frequency normalized to the high rate, so that
$z=e^{j\theta}=e^{j\omega/M}$; the pass band is $|\omega|\le\omega_p<\pi$. The
normalized magnitude response is
\begin{align}
\frac{|H_{N,M}(e^{j\omega/M})|}{M^{N}}
&=\left|\frac{\sin(\omega/2)}{M\sin(\omega/2M)}\right|^{N}\nonumber\\
&= S_N(\omega)\,\bigl[1+\varepsilon^{\rm comb}_{N,M}(\omega)\bigr],
\label{eq:combresp}
\end{align}
where
\begin{equation}
S_N(\omega)=\left|\frac{\sin(\omega/2)}{\omega/2}\right|^{N}
\end{equation}
is the response of the continuous-time $N$-fold moving average, which does not
depend on $M$, and
\begin{equation}
\varepsilon^{\rm comb}_{N,M}(\omega)=\frac{N}{24}\Bigl(\frac{\omega}{M}\Bigr)^{2}
+O(M^{-4})
\label{eq:epscomb}
\end{equation}
is the \emph{pass-band deviation} of the plain comb. It is positive: for
finite $M$ the comb droops less than $S_N(\omega)$, which is the worst case,
reached for $M\to\infty$. A compensator designed
for $S_N(\omega)$ is rate independent; the deviation is what makes the droop,
and hence the compensator, depend on $M$. The symbols
$\varepsilon^{\rm comb}$, $\varepsilon^{D}$ and $\varepsilon^{E}$ denote the
deviation of the plain comb, of the comb with the three-tap filter $D_N$, and
of the comb with the Eulerian filter $E_N$, respectively.

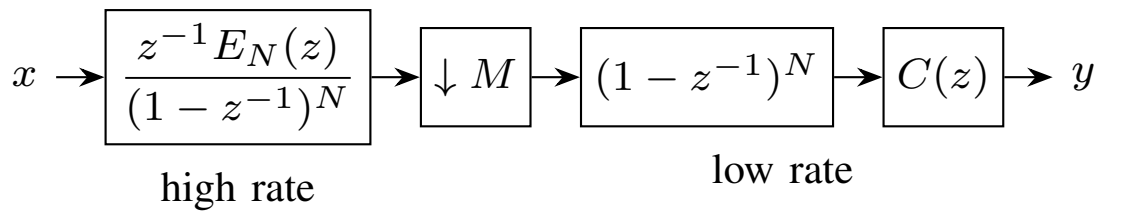
\begin{figure}[t]
\centering
\begin{tikzpicture}[>=Stealth,font=\footnotesize,
  blk/.style={draw,minimum height=6.5mm,inner sep=2.5pt},node distance=3.2mm]
\node (x) {$x$};
\node[blk,right=of x] (i) {$\dfrac{z^{-1}E_N(z)}{(1-z^{-1})^{N}}$};
\node[blk,right=of i] (d) {$\downarrow M$};
\node[blk,right=of d] (c) {$(1-z^{-1})^{N}$};
\node[blk,right=of c] (k) {$C(z)$};
\node[right=of k] (y) {$y$};
\draw[->] (x)--(i); \draw[->] (i)--(d); \draw[->] (d)--(c); \draw[->] (c)--(k); \draw[->] (k)--(y);
\node[below=0.5mm of i] {high rate};
\node[below=0.5mm of c,xshift=5mm] {low rate};
\end{tikzpicture}
\caption{Derated comb decimator. The derating filter
$E_N$ sits in the integrator section at the high rate; the droop compensator
$C(z)$ sits at the low rate. With $E_N$ equal to the Euler--Frobenius
polynomial, the coefficients of both depend on $N$ only.}
\label{fig:blk}
\end{figure}

The structure considered in this paper is shown in Fig.~\ref{fig:blk}. The
comb \eqref{eq:comb} is realized in the usual way \cite{hogenauer}, with the
$N$ integrators at the high rate and the $N$ differentiators at the low rate;
a fixed FIR filter is added to the integrator section, and the droop
compensator $C(z)$ remains at the low rate.

\subsection{Three-tap derating filter}

The method of \cite{lee} cascades with \eqref{eq:comb} the filter
\begin{equation}
D_N(z)=\frac{1+b_N z^{-1}+z^{-2}}{2+b_N},\qquad b_N=\frac{24}{N}-2 ,
\label{eq:DN}
\end{equation}
whose zero-phase response $1-\tfrac{N}{12}(1-\cos\theta)$ cancels the $M^{-2}$ term of \eqref{eq:epscomb}. Expanding
one order further (Appendix~\ref{app:a}) gives the deviation of the derated comb,
\begin{equation}
\varepsilon^{D}_{N,M}(\omega)=\frac{N(22-5N)}{5760}
\Bigl(\frac{\omega}{M}\Bigr)^{4}+O(M^{-6}).
\label{eq:epsD}
\end{equation}
In \cite{lee} the method is valid for $N<12$, the range in which $b_N>0$ and
the stop band of the comb is not degraded, and the response decreases
monotonically for $N\le6$, where $b_N\ge2$ and $D_N$ has no zeros on the unit
circle other than at the high-rate Nyquist frequency.

\subsection{Eulerian numbers}

The Eulerian number $\eul{n}{j}$ counts the permutations of $\{1,\dots,n\}$
with exactly $j$ descents \cite{gkp,petersen},
\begin{equation}
\eul{n}{j}=\sum_{i=0}^{j}(-1)^{i}\binom{n+1}{i}(j+1-i)^{n},
\quad 0\le j\le n-1 .
\end{equation}
The Eulerian polynomial is $A_n(x)=\sum_{j=0}^{n-1}\eul{n}{j}x^{j}$, with
$A_0(x)=1$; thus $A_1=1$, $A_2=1+x$, $A_3=1+4x+x^2$,
$A_4=1+11x+11x^2+x^3$. Euler's identity reads
\begin{equation}
\sum_{k\ge 0}k^{n}x^{k}=\frac{x\,A_n(x)}{(1-x)^{n+1}},\qquad n\ge1 .
\label{eq:euler}
\end{equation}
Expanding $(1-x)^{-(n+1)}$ and using the symmetry
$\eul{n}{j}=\eul{n}{n-1-j}$, the same statement reads, coefficient by
coefficient,
\begin{equation*}
k^{n}=\sum_{j=0}^{n-1}\eul{n}{j}\binom{k+j}{n},
\end{equation*}
which is Worpitzky's identity \cite{worpitzky,gkp}. It is the time-domain form
of \eqref{eq:euler}: $\binom{k+n}{n}$ is the impulse response of $n+1$
cascaded integrators, so $k^{n}$ is the sum of $n$ shifted copies of that
response, weighted by the Eulerian numbers. Comparing the coefficients of
$k^{n}$ on both sides gives $\sum_j\eul{n}{j}=n!$, that is $A_n(1)=n!$; this
is the value at $z=1$ ($\omega=0$), and it fixes the normalization of the
filter defined below. The zeros of $A_n$
are real, negative and simple \cite{frobenius,schoenberg}. In spline theory
$A_n$ is known, up to normalization, as the Euler--Frobenius polynomial
\cite{schoenberg,he}.

\section{Exact Derating Filter}

Let $\beta^{n}(t)$ denote the causal B-spline of degree $n$ with knots at the
integers, supported on $[0,n+1]$,
\begin{equation}
\beta^{n}(t)=\frac{1}{n!}\sum_{i=0}^{n+1}(-1)^{i}\binom{n+1}{i}(t-i)_{+}^{n},
\label{eq:bspline}
\end{equation}
that is, the $(n+1)$-fold convolution of the unit box, whose Fourier transform
has magnitude $S_{n+1}$. Define the \emph{Eulerian derating filter} of order
$N\ge2$ as
\begin{equation}
E_N(z)=\frac{A_{N-1}(z^{-1})}{(N-1)!}
=\frac{1}{(N-1)!}\sum_{j=0}^{N-2}\eul{N-1}{j}z^{-j}.
\label{eq:EN}
\end{equation}

\begin{theorem}\label{thm:main}
For every $N\ge2$ and $M\ge1$, the impulse response $g_{N,M}[k]$ of
\begin{equation}
G_{N,M}(z)=z^{-1}M^{-N}H_{N,M}(z)\,E_N(z)
\label{eq:G}
\end{equation}
is
\begin{equation}
g_{N,M}[k]=\frac{1}{M}\,\beta^{N-1}\!\Bigl(\frac{k}{M}\Bigr),\qquad k\ge0 .
\label{eq:main}
\end{equation}
\end{theorem}

\begin{proof}
Put $n=N-1$. Sampling \eqref{eq:bspline} at $t=k/M$ gives
\begin{equation*}
\beta^{n}\!\Bigl(\frac kM\Bigr)=\frac{1}{n!\,M^{n}}\sum_{i=0}^{n+1}(-1)^{i}
\binom{n+1}{i}\,(k-iM)_{+}^{n}.
\end{equation*}
The $z$-transform of $(k-iM)_{+}^{n}$ is $z^{-iM}\sum_{k\ge0}k^{n}z^{-k}$,
and $\sum_{i}(-1)^{i}\binom{n+1}{i}z^{-iM}=(1-z^{-M})^{n+1}$, so
\begin{align*}
\sum_{k\ge0}\beta^{n}\!\Bigl(\frac{k}{M}\Bigr)z^{-k}
&=\frac{(1-z^{-M})^{n+1}}{n!\,M^{n}}\sum_{k\ge0}k^{n}z^{-k}\\
&=\frac{(1-z^{-M})^{n+1}}{n!\,M^{n}}\cdot
\frac{z^{-1}A_n(z^{-1})}{(1-z^{-1})^{n+1}},
\end{align*}
where the last step is Euler's identity \eqref{eq:euler} with $x=z^{-1}$.
Since $n+1=N$, the right-hand side is
$M\cdot z^{-1}M^{-N}H_{N,M}(z)\,A_{N-1}(z^{-1})/(N-1)!=M\,G_{N,M}(z)$.
\end{proof}

Theorem~\ref{thm:main} says that the comb followed by $E_N$ is no longer an
approximation of the B-spline kernel: it \emph{is} that kernel, sampled on the
high-rate grid. The decimation factor enters only as the sampling density. The one-sample
delay in \eqref{eq:G} is the delay of one pipelined integrator and costs
nothing; it places the origin of the sampling grid on the knot $t=0$ of the
spline. The grids $\{k/M\}$ for different $M$ then all contain the knots
$t=0,1,2,\dots$, and at these common instants the impulse responses of all
rates take the same values $\beta^{N-1}(p)$, which are the Eulerian numbers
divided by $(N-1)!$. Before normalization the knot values are
$M^{N-1}\eul{N-1}{p-1}$: integers, exactly representable in fixed-point
arithmetic for every $M$, and only $N-1$ of them are nonzero.

\begin{remark}[Relation to known results]
With $n=N-1$ and $m=M$, \eqref{eq:main} is equivalent to the identity
$U^{n}_{m}(z)=m^{-n}U^{n}_{1}(z)\,(U^{0}_{m}(z))^{n+1}$ of
\cite[Eq.~(B-3)]{unser93a} (see also \cite{unser91}), in which $U_1^n$ is the
$z$-transform of the integer-sampled B-spline. In \cite{unser93a} the identity
is an intermediate step towards centered (zero-phase) kernels, as needed for
images. Centering forces a case distinction on the parity of $n$ and $m$,
with two different correction kernels and half-sample shifts, and the
integer-sampled kernels of even degree are then no longer Eulerian. The
causal form used here needs no case distinction: all rates and all orders are
aligned to the single common origin $t=0$, and the coefficients are the
Eulerian numbers throughout. No claim of novelty is made for the identity
itself.
\end{remark}

\begin{corollary}\label{cor:prop}
For every $N\ge2$: (a) $E_N$ has $N-1$ symmetric positive integer coefficients
divided by $(N-1)!$, hence exact linear phase with a delay of $(N-2)/2$ high-rate
samples and unit DC gain; (b) $|E_N(e^{j\theta})|\le1$ and is monotonically
decreasing on $[0,\pi]$, so no stop-band attenuation of the comb is lost and
the pass band stays monotonic; (c) $G_{N,M}$ has a group delay of exactly $NM/2$ high-rate samples, that
is, $N/2$ low-rate samples for every $M$ and for even and odd $N$ alike.
\end{corollary}

\begin{proof}
(a) follows from the symmetry and the row sum $(N-1)!$ of the Eulerian
numbers. For (b),
since all zeros $r_i$ of $A_{N-1}$ are real and negative, each factor
$|e^{j\theta}-r_i|$ decreases on $[0,\pi]$ and is largest at $\theta=0$. (c)
follows from \eqref{eq:main}, since $\beta^{N-1}$ is symmetric about $t=N/2$;
equivalently, the delays $N(M-1)/2$ of the comb, $(N-2)/2$ of $E_N$ and $1$ of
\eqref{eq:G} add up to $NM/2$.
\end{proof}

Part (b) holds for all $N$, so neither the bound $N<12$ nor the monotonicity
bound $N\le6$ of the three-tap design arises. Part (c) is the alignment property: when the rate is switched,
or when several rates run side by side, the delay measured in low-rate samples
does not move, and no parity-dependent half-sample correction is needed. For odd $N$ the polynomial $A_{N-1}$ has a zero at $-1$, which
adds a transmission zero at the high-rate Nyquist frequency.

\section{Derivation from Rate Invariance}\label{sec:ri}

Theorem~\ref{thm:main} starts from the B-spline. The filter can also be
obtained without any reference to continuous-time splines, from a single
requirement on the discrete structure: \emph{doubling the rate must not change
the values of the response at the instants common to both rates.} This is the
route by which $E_N$ was originally found, and it shows that the filter is not
one choice among several but the only one.

Before the comb section, the structure consists of $N$ integrators and a
correction filter $P(z)$. Let $y[k]$ be the impulse response of
$P(z)/(1-z^{-1})^{N}$, and put $n=N-1$.

\begin{theorem}\label{thm:ri}
Let $P(z)$ be a causal FIR filter and $n=N-1\ge1$. The response $y[k]$
satisfies the rate-invariance condition
\begin{equation}
y[2k]=2^{n}\,y[k]\qquad\text{for all }k\ge0
\label{eq:ri}
\end{equation}
if and only if $P(z)=c\,z^{-1}A_{n}(z^{-1})$ for a constant $c$, that is, if
and only if $P$ is the Eulerian filter with the one-sample delay of
\eqref{eq:G}. In that case $y[k]=c\,k^{n}$, and $y[rk]=r^{n}y[k]$ for every
integer $r\ge1$.
\end{theorem}

\begin{proof}
If $P=c\,z^{-1}A_n(z^{-1})$, then $y[k]=c\,k^{n}$ by \eqref{eq:euler}, which
satisfies \eqref{eq:ri}. Conversely, for $k$ larger than the degree of $P$ the
response of $N$ integrators is a polynomial $q(k)$ of degree at most $n$.
Condition \eqref{eq:ri} for infinitely many $k$ gives $q(2k)=2^{n}q(k)$
identically, so the coefficient $q_i$ of $k^{i}$ obeys $2^{i}q_i=2^{n}q_i$ and
$q(k)=c\,k^{n}$. For any $k\ge1$ choose $j$ with $2^{j}k$ beyond the degree of
$P$; then $y[k]=2^{-jn}y[2^{j}k]=c\,k^{n}$, and $y[0]=2^{n}y[0]$ gives
$y[0]=0$. Hence $y[k]=c\,k^{n}$ for all $k$, and
$P(z)=(1-z^{-1})^{N}\sum_k c\,k^{n}z^{-k}=c\,z^{-1}A_n(z^{-1})$.
\end{proof}

The factor $2^{n}$ in \eqref{eq:ri} is a pure gain and is removed by the
normalization $M^{-N}$ of \eqref{eq:G}. The comb section
$(1-z^{-M})^{N}$ then turns the rate-invariant response $k^{n}$ into the
B-spline samples of Theorem~\ref{thm:main}, and \eqref{eq:ri} becomes
\begin{equation}
g_{N,2M}[2k]=\tfrac12\,g_{N,M}[k],
\end{equation}
i.e.\ the kernel at rate $2M$, read at every second sample, is the kernel at
rate $M$.

Rate invariance thus holds \emph{exactly in the time domain}: the values of
the impulse response at the common instants and the group delay do not depend
on $M$. In the frequency domain it holds up to the aliasing term derived in
the next section, which is why the time-domain statement is taken as the
defining one.

\begin{remark}[Original formulation and construction rule]
The requirement was first stated for the recursion itself rather than for the
impulse response. Let $y_i[k]$, $i=0,\dots,n$, be the outputs of the stages
of the structure, with $y_i[k]=k^{i}$ in the absence of input. By the binomial
theorem,
\begin{equation}
y_n[k]=\sum_{i=0}^{n}\binom{n}{i}\,y_{i}[k-1],
\end{equation}
that is, $\mathbf y[k]=\mathbf P\,\mathbf y[k-1]$ with $\mathbf P$ the
lower-triangular Pascal matrix. Rate invariance is then the statement that
two steps at the doubled rate equal one step at the original rate once the
$i$th stage is rescaled by $2^{i}$,
\begin{equation*}
y_n[k]=\sum_{i=0}^{n}\binom{n}{i}\,2^{\,n-i}\,y_{i}[k-2],
\quad\text{i.e.}\quad \mathbf P^{2}=\mathbf D\mathbf P\mathbf D^{-1},
\end{equation*}
with $\mathbf D=\mathrm{diag}(1,2,\dots,2^{n})$. It follows from the
elementary identity $\sum_{i=j}^{n}\binom{n}{i}\binom{i}{j}=\binom{n}{j}2^{n-j}$,
and holds for every integer ratio $r$ with $\mathrm{diag}(r^{i})$ in place of
$\mathbf D$. Conversely, writing a unit lower-triangular update matrix as
$\mathbf A=\exp\mathbf L$, the condition
$\mathbf A^{2}=\mathbf D\mathbf A\mathbf D^{-1}$ forces
$2\mathbf L=\mathbf D\mathbf L\mathbf D^{-1}$, so $\mathbf L$ is confined to
the first subdiagonal and $\mathbf A=\mathbf S\mathbf P\mathbf S^{-1}$ for a
diagonal $\mathbf S$ whenever that subdiagonal has no zero entry: up to the
gains of the individual stages, the Pascal recursion is the only
rate-invariant one. This is the state-space form of Theorem~\ref{thm:ri}.

The rate-invariant response of order $n$ is thus built recursively from those
of lower order with binomial weights, in the same way as the response of a
plain integrator chain of order $n$ is built from those of lower order with
unit weights. Dividing the two recursions in the $z$-domain yields a recursion
for $A_n$ and reproduces $1$; $1,1$; $1,4,1$; $1,11,11,1$; \dots{} without the
closed form. This route uses elementary arithmetic only; it is how the filters
were first derived, independently of spline theory and of the Eulerian
numbers.
\end{remark}

\section{Residual Rate Dependency}

\begin{proposition}\label{prop:eps}
For $N\ge2$, with $z=e^{j\omega/M}$,
\begin{equation}
G_{N,M}(z)=e^{-j\omega N/2}\,
\Bigl(\frac{\sin(\omega/2)}{\omega/2}\Bigr)^{N}
\bigl[1+\varepsilon^{E}_{N,M}(\omega)\bigr],
\end{equation}
where the pass-band deviation is the real quantity
\begin{equation}
\varepsilon^{E}_{N,M}(\omega)=\sum_{l\ne0}
\Bigl(\frac{\omega}{\omega+2\pi lM}\Bigr)^{N}
\label{eq:epsE}
\end{equation}
(summed symmetrically for $N=2$). Its leading term is
\begin{equation}
\varepsilon^{E}_{N,M}(\omega)\simeq
\begin{cases}
\dfrac{|B_N|}{N!}\Bigl(\dfrac{\omega}{M}\Bigr)^{N}, & N \text{ even},\\[2.2ex]
-\dfrac{N\,|B_{N+1}|}{(N+1)!}\Bigl(\dfrac{\omega}{M}\Bigr)^{N+1}, & N \text{ odd},
\end{cases}
\label{eq:lead}
\end{equation}
where $B_k$ are the Bernoulli numbers.
\end{proposition}

\begin{proof}
By Theorem~\ref{thm:main}, $g_{N,M}[k]$ are the integer samples of
$f(t)=M^{-1}\beta^{N-1}(t/M)$, whose Fourier transform is
$F(\Omega)=\hat\beta(M\Omega)$ with
$\hat\beta(\Omega)=e^{-j\Omega N/2}\,(\sin(\Omega/2)/(\Omega/2))^{N}$. The
Poisson summation formula gives
$G_{N,M}(e^{j\theta})=\sum_{l}F(\theta+2\pi l)$. Writing $\omega=M\theta$ and
using $\sin(\omega/2+\pi lM)=(-1)^{lM}\sin(\omega/2)$, the factors
$(-1)^{lMN}$ from the sine and from the phase cancel, and every term equals
$e^{-j\omega N/2}\sin^{N}(\omega/2)\,(\omega/2+\pi lM)^{-N}$.
Dividing by the $l=0$ term gives \eqref{eq:epsE}. Expanding each term in
$u_l=\omega/(2\pi lM)$ as $u_l^{N}(1-Nu_l+\cdots)$, the odd powers cancel
between $l$ and $-l$, and $2\zeta(2k)=(2\pi)^{2k}|B_{2k}|/(2k)!$ gives
\eqref{eq:lead}.
\end{proof}

The deviation \eqref{eq:epsE} is the aliasing of the tails of the
continuous-time response $S_N$ around the multiples of the high rate. It is
the only rate dependency left.

Table~\ref{tab:coef} lists the coefficients of $E_N$ and the leading terms of
\eqref{eq:lead} and \eqref{eq:epsD}. Three cases deserve comment.

\emph{$N=2$.} Here $E_2=1$: the second-order comb is already the sampled
linear B-spline, and $\varepsilon^{E}=\varepsilon^{\rm comb}$. Exactness in the
time domain does not reduce the rate dependency of the pass band here, and the
three-tap filter $D_2$ of \cite{lee}, which matches $S_2$ rather than its
aliased sum, is the better choice in that respect.

\emph{$N=4$.} $E_4(z)=(1+4z^{-1}+z^{-2})/6=D_4(z)$. The three-tap filter of
\cite{lee} is exact, and \eqref{eq:epsD} and \eqref{eq:lead} agree
($1/720$).

\emph{$N=3$.} $E_3(z)=(1+z^{-1})/2$ has only two taps. The filter
$D_3(z)=(1+6z^{-1}+z^{-2})/8$ is not $E_3$, but it is also exact: it yields
the quadratic B-spline sampled on a grid offset by half a high-rate sample,
$(k+\tfrac12)/M$ (verified in exact arithmetic). It is the member with
offset $\tfrac12$ of the generalized Euler--Frobenius family
\cite{schoenberg}, obtained by replacing $k^{n}$ by $(k+\tau)^{n}$ in
\eqref{eq:euler}. The two-tap filter $E_3$ is the simplest non-trivial member
of the family, and $D_6(z)=E_3(z)^{2}$.

For $N\ge5$ the two filters differ. A symmetric three-tap filter with unit DC
gain has one free parameter, and $D_N$ spends it on the second moment $N/12$,
which is the variance of $\beta^{N-1}$; $D_N$ is thus the second-moment
approximation of $E_N$. Its deviation stays at $M^{-4}$, whereas that of $E_N$
falls to $M^{-6}$ for $N=5,6$ and $M^{-8}$ for $N=7,8$.

\begin{table*}[t]
\centering
\caption{Eulerian derating filter $E_N$, word-length increase, and leading
term of the pass-band deviation ($\theta=\omega/M$). The word-length increase
of $D_N$ is $W_b$ of \cite{lee}.}
\label{tab:coef}
\begin{tabular}{@{}clrccll@{}}
\toprule
 & & & \multicolumn{2}{c}{extra bits} & \multicolumn{2}{c}{leading term of the deviation}\\
\cmidrule(lr){4-5}\cmidrule(l){6-7}
$N$ & coefficients of $(N-1)!\,E_N(z)$ & $(N-1)!$ & $E_N$ & $D_N$ \cite{lee} &
$\varepsilon^{E}$ (Eulerian filter) & $\varepsilon^{D}$ (three-tap filter)\\
\midrule
2 & 1 & 1 & 0 & 4 & $\theta^{2}/12$ & $\theta^{4}/240$\\
3 & 1, 1 & 2 & 1 & 3 & $-\theta^{4}/240$ & $7\,\theta^{4}/1920$\\
4 & 1, 4, 1 & 6 & 3 & 3 & $\theta^{4}/720$ & $\theta^{4}/720$\\
5 & 1, 11, 11, 1 & 24 & 5 & 5 & $-\theta^{6}/6048$ & $-\theta^{4}/384$\\
6 & 1, 26, 66, 26, 1 & 120 & 7 & 2 & $\theta^{6}/30240$ & $-\theta^{4}/120$\\
7 & 1, 57, 302, 302, 57, 1 & 720 & 10 & 5 & $-\theta^{8}/172800$ & $-91\,\theta^{4}/5760$\\
8 & 1, 120, 1191, 2416, 1191, 120, 1 & 5040 & 13 & 2 & $\theta^{8}/1209600$ & $-\theta^{4}/40$\\
\bottomrule
\end{tabular}
\end{table*}

\section{Rate-Invariant Compensation}\label{sec:comp}

Every term of \eqref{eq:epsE} carries the factor $\omega^{N}$, for every $M$.
This gives the main practical statement.

\begin{corollary}\label{cor:taylor}
With the Eulerian derating filter, the Taylor expansion about $\omega=0$ of
the pass-band response, referred to the low rate, coincides with that of
$S_N(\omega)$ up to and including the term $\omega^{N-1}$ ($N$ even) or
$\omega^{N}$ ($N$ odd), for every $M$. Consequently the coefficients of any
compensator designed from these Taylor terms are functions of $N$ alone.
\end{corollary}

Maximally flat compensators \cite{fernandez,molnar} are designed exactly in
this way. Closed-form coefficients are given in \cite{fernandez}; for the plain
comb they are functions of both $N$ and $M$. Consider the symmetric three-tap
compensator $C(z)=c_0+(1-2c_0)z^{-1}+c_0z^{-2}$ at the low rate. Matching the
$\omega^{2}$ term of the plain comb \eqref{eq:combresp} gives the
rate-dependent coefficient \cite[Eq.~(12)]{lee}
\begin{equation}
c_0=-\frac{N}{24}\Bigl(1-\frac{1}{M^{2}}\Bigr),
\label{eq:c0M}
\end{equation}
whereas after derating the target is $1/S_N$ and
\begin{equation}
c_0=-\frac{N}{24}
\label{eq:c0}
\end{equation}
for every $M$. For the three-tap compensator this already holds with $D_N$
\cite{lee}, whose deviation is $O(\omega^{4})$. The five-tap maximally flat
compensator, with taps $(c_b,c_a,1-2c_a-2c_b,c_a,c_b)$, must also match the
$\omega^{4}$ term; for the target $1/S_N$ this gives (Appendix~\ref{app:b})
\begin{equation}
c_a=-\frac{N(5N+82)}{1440},\qquad c_b=\frac{N(5N+22)}{5760}.
\label{eq:c5}
\end{equation}
These coefficients are exact for every $M$ when the deviation is
$O(\omega^{6})$, which holds for $E_N$ with $N\ge5$. With $D_N$ they hold
approximately, as noted in \cite{lee} for the wide-band case, because the
$\omega^{4}$ term \eqref{eq:epsD} still depends on $M$, through $M^{-4}$. In general a
maximally flat compensator with $2K+1$ taps has rate-independent closed-form
coefficients if $2K<N$ ($N$ even) or $2K\le N$ ($N$ odd).

\begin{remark}[Interpolators and recursive realizations]
A comb interpolator uses the same transfer function \eqref{eq:G} after
up-sampling by $M$, so the results of this paper carry over, with the
compensator placed before the interpolator at the low rate. By
Theorem~\ref{thm:main} its output for the input $x[m]$ is
\begin{equation*}
y[k]=\sum_{m}x[m]\,\beta^{N-1}\!\Bigl(\frac kM-m\Bigr).
\end{equation*}
At the common instants $k=pM$ this becomes
\begin{equation*}
y[pM]=\sum_{m}x[m]\,\beta^{N-1}(p-m),
\end{equation*}
the input filtered by $z^{-1}E_N(z)$ \emph{at the low rate}, the same for
every $M$. Two facts are worth recording
here. First, the overall filter is FIR although it is realized recursively
with integrators, so its response is determined by finitely many inputs.
Second, at the common instants the output, which is the state of the last
integrator, is an exact integer combination (before normalization) of only the
$N-1$ most recent inputs, with Eulerian weights that do not depend on $M$.
The state of a recursive realization is therefore known exactly, at every
common instant, from the last $N-1$ inputs alone. Realizations that exploit
this, and their behaviour in finite word length, are the subject of a
separate paper.
\end{remark}

\section{Worked Example: \texorpdfstring{$N=5$}{N=5}}

For a fifth-order comb the filters of Fig.~\ref{fig:blk} are, for every $M$,
\begin{align*}
E_5(z)&=\tfrac{1}{24}\bigl(1+11z^{-1}+11z^{-2}+z^{-3}\bigr),\\
C(z)&=\tfrac{47}{1152}\,(1+z^{-4})-\tfrac{107}{288}\,(z^{-1}+z^{-3})
+\tfrac{319}{192}\,z^{-2},
\end{align*}
the latter being \eqref{eq:c5} with $N=5$. The integrator section with $E_5$
can be realized as the tapped chain of Section~\ref{sec:impl} with weights
$1,14,36,24$ on stages $2$ to $5$. The register width grows by
$\lceil\log_2 24\rceil=5$ bits, and the group delay of the derated comb is
$2.5$ low-rate samples. For comparison, the three-tap filter of \cite{lee} is
$D_5(z)=(1+2.8z^{-1}+z^{-2})/4.8$.

At $\omega=\pi/4$ the deviation from the rate-independent response $S_5$ is,
for $M=4$, $8$ and $16$: $8.1\times10^{-3}$, $2.0\times10^{-3}$,
$5.0\times10^{-4}$ for the comb alone; $3.9\times10^{-6}$, $2.4\times10^{-7}$,
$1.5\times10^{-8}$ in magnitude with $D_5$; and $9.5\times10^{-9}$,
$1.5\times10^{-10}$, $2.3\times10^{-12}$ in magnitude with $E_5$. The same
$C(z)$ is therefore valid for all three rates to better than $10^{-8}$, where
the plain comb would need three different compensators.

\section{Implementation}\label{sec:impl}

\subsection{Direct form}

$E_N$ can be placed before or after the integrators, at the high rate, like
$D_N$ in \cite{lee}. Its coefficients are small integers that are realized by
shifts and additions, and by symmetry only $\lceil(N-1)/2\rceil$ of them are
distinct. The division by $(N-1)!$ is absorbed in the output scaling. Since the
cascade is an FIR filter with non-negative integer coefficients and DC gain
$(N-1)!\,M^{N}$, the modular (wrap-around) arithmetic of \cite{hogenauer}
remains valid when the register width is increased by
$\lceil\log_2 (N-1)!\rceil$ bits (Table~\ref{tab:coef})
over the $\lceil N\log_2M\rceil$ bits of growth of the comb. Table~\ref{tab:coef} also lists the
corresponding increase $W_b$ for $D_N$ from \cite{lee}. For $N\le5$ the
Eulerian filter needs no more bits than the three-tap filter (1 against 3 for
$N=3$, equal for $N=4,5$); from $N=6$ on it needs more (7 against 2 for
$N=6$). This is the cost of exactness.

\subsection{Tapped integrator chain}

Expanding $k^{n}$ in binomial coefficients gives the classical alternative to
\eqref{eq:euler} \cite{gkp},
\begin{equation}
\frac{x\,A_n(x)}{(1-x)^{n+1}}=\sum_{j=1}^{n}j!\stir{n}{j}
\frac{x^{j}}{(1-x)^{j+1}},
\label{eq:stirling}
\end{equation}
where $\stir{n}{j}$ are the Stirling numbers of the second kind. Let $I_0(z)=1/(1-z^{-1})$ be an integrator without delay and
$I(z)=z^{-1}/(1-z^{-1})$ a pipelined one. With $x=z^{-1}$,
$x^{j}/(1-x)^{j+1}=I_0(z)I(z)^{j}$, and \eqref{eq:stirling} becomes
\begin{equation}
\frac{z^{-1}E_N(z)}{(1-z^{-1})^{N}}=\frac{1}{(N-1)!}\sum_{j=1}^{N-1}
j!\stir{N-1}{j}\,I_0(z)\,I(z)^{j}.
\end{equation}
The integrator section of \eqref{eq:G}, including $E_N$ and the one-sample
delay, is therefore a weighted sum of the outputs of stages $2,\dots,N$ of an
ordinary integrator chain whose first stage is undelayed and whose other
stages are pipelined. No delay line is added, and the number of taps, $N-1$, equals the length of
$E_N$; the weights are listed in Table~\ref{tab:stir}.
The transfer function is unchanged, so the word-length statement above
applies.

\begin{table}[!t]
\centering
\caption{Weights $j!\stir{N-1}{j}$ of the tapped integrator chain
(stage $j+1$, $j=1,\dots,N-1$).}
\label{tab:stir}
\begin{tabular}{@{}cl@{}}
\toprule
$N$ & weights\\
\midrule
3 & 1, 2\\
4 & 1, 6, 6\\
5 & 1, 14, 36, 24\\
6 & 1, 30, 150, 240, 120\\
7 & 1, 62, 540, 1560, 1800, 720\\
\bottomrule
\end{tabular}
\end{table}

\section{Numerical Results}

Theorem~\ref{thm:main} was checked in exact rational arithmetic for
$N=2,\dots,9$ and $M=1,\dots,8$: the convolution of the comb impulse response
with the Eulerian coefficients, delayed by one sample, equals
$(N-1)!\,M^{N-1}\beta^{N-1}(k/M)$ for every $k$. Python scripts that
reproduce this check, all other checks, and every table and figure of this
paper are provided as ancillary files.

\begin{figure*}[t]
\centering
\includegraphics[width=0.88\textwidth]{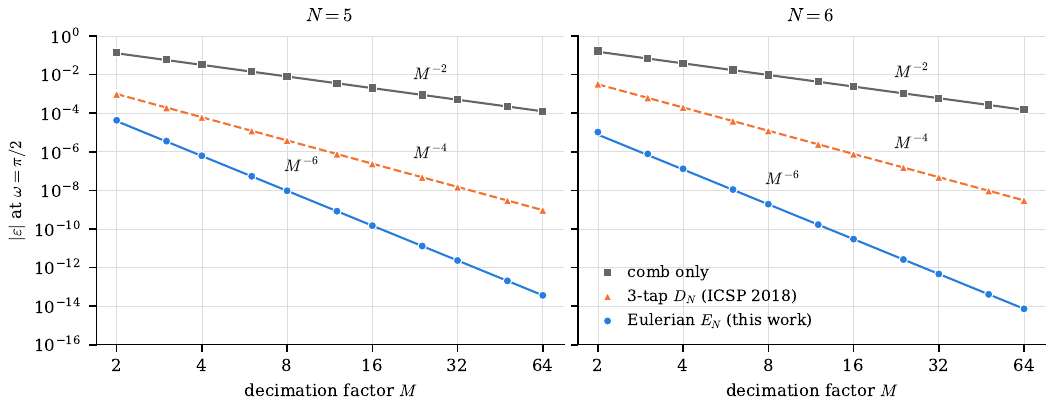}
\caption{Pass-band deviation $|\varepsilon|$ from the rate-independent response
$S_N(\omega)$ at $\omega=\pi/2$ versus decimation factor. Markers: computed
from the frequency responses. Lines: leading terms \eqref{eq:epscomb},
\eqref{eq:epsD}, \eqref{eq:lead}.}
\label{fig:res}
\end{figure*}

Fig.~\ref{fig:res} shows the magnitude of the pass-band deviation at
$\omega=\pi/2$ (the pass-band edge when a further decimation by two follows)
for $N=5$ and $N=6$, computed directly from the frequency responses in
extended precision (markers) together with the leading terms
\eqref{eq:epscomb}, \eqref{eq:epsD} and \eqref{eq:lead} (lines). The slopes
$M^{-2}$, $M^{-4}$ and $M^{-6}$ are confirmed. At $M=8$ and $N=6$ the
deviation is $9.7\times10^{-3}$ for the comb, $1.2\times10^{-5}$ with the
three-tap filter and $1.9\times10^{-9}$ with the Eulerian filter; at $M=32$
the three values are $6.0\times10^{-4}$, $4.8\times10^{-8}$ and
$4.6\times10^{-13}$.

For a compensator designed once for $S_N(\omega)$, these numbers are the
relative gain error caused by changing the rate. Table~\ref{tab:spread} shows
the effect on a complete chain: the fixed five-tap compensator \eqref{eq:c5}
for $N=6$ is applied without change for $M=4,\dots,256$, and the spread of the
compensated gain over $M$ is recorded at two band edges. The spread is the
rate dependency seen by the user; the flatness error of the five-tap
compensator itself ($0.02$~dB at $\omega=\pi/4$) is common to all three
columns and does not depend on $M$ once the comb is derated.

\begin{figure*}[t]
\centering
\includegraphics[width=0.88\textwidth]{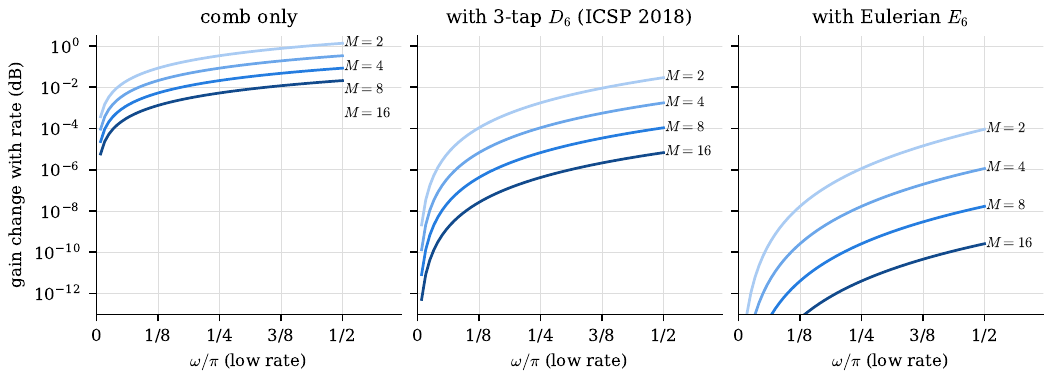}
\caption{Change of pass-band gain with the decimation factor, $N=6$:
$|20\log_{10}(1+\varepsilon)|$ versus low-rate frequency, for $M=2,4,8,16$.
A fixed low-rate compensator leaves exactly this error when the rate is
changed.}
\label{fig:rate}
\end{figure*}

Fig.~\ref{fig:rate} shows the same effect over the whole pass band for
$N=6$: the change of gain, in dB, with respect to the limit $M\to\infty$. For
the plain comb the curves for different $M$ lie between $10^{-2}$ and $1$~dB
at the band edge; with $D_6$ they fall to $10^{-5}$--$10^{-1}$~dB, and with
$E_6$ to $10^{-10}$--$10^{-4}$~dB, the largest value occurring for $M=2$.

\emph{Stop band.} Because $|E_N|\le1$, the attenuation in the folding bands
cannot decrease. For $N=6$ and a pass band $|\omega|\le\pi/4$, the worst-case
alias attenuation is $101.7$~dB for the comb and $102.8$~dB with $E_6$ at
$M=8$ ($102.5$ and $102.8$~dB at $M=16$); for $|\omega|\le\pi/2$ the
corresponding values are $62.0$ and $62.7$~dB. The change is below $1.1$~dB
and always an improvement.

\begin{table}[t]
\centering
\caption{Spread over $M\in\{4,8,16,32,64,256\}$ of the compensated gain in
dB, $N=6$, with the fixed compensator \eqref{eq:c5}.}
\label{tab:spread}
\begin{tabular}{@{}lccc@{}}
\toprule
band edge $\omega$ & comb only & with $D_6$ \cite{lee} & with $E_6$\\
\midrule
$\pi/4$ & $8.4\times10^{-2}$ & $1.1\times10^{-4}$ & $1.7\times10^{-8}$\\
$\pi/2$ & $3.4\times10^{-1}$ & $1.7\times10^{-3}$ & $1.1\times10^{-6}$\\
\bottomrule
\end{tabular}
\end{table}

\section{Related Work}

The identity behind Theorem~\ref{thm:main} belongs to cardinal spline theory
\cite{schoenberg}; its signal-processing form is due to Unser \emph{et al.}
\cite{unser91,unser93a,unser93b}, who use cascaded moving averages as a fast
algorithm for image interpolation with a fixed, small expansion factor. The
relation between Eulerian polynomials and B-splines is surveyed in \cite{he},
and the same polynomials govern the sampling zeros of sampled-data systems
\cite{weller}. None of these works addresses comb decimators, the alignment of several
rates to a common delay, or the design of rate-independent droop
compensators.

Sample-rate converters that combine comb filters with polynomial-based
(Farrow-type) interpolation \cite{hentschel,babic02,babic05} obtain arbitrary
conversion ratios, at the price of multipliers or of coefficients that must be
updated when the ratio changes. A polynomial interpolator can be made
multiplier-free by realizing it with integrators, that is, as a comb; but the
piecewise-polynomial coefficients of the plain comb depend on $M$ (already for
$N=3$ every segment contains terms in $1/M$), and the expressions become
involved as $N$ grows; its spectrum then changes with the rate, which the
spectrum of a true polynomial interpolator does not. By
Theorem~\ref{thm:main} the derated comb has the fixed polynomial pieces of
the B-spline instead: it is a multiplier-free realization of the B-spline
interpolator at integer ratios, and it recovers the rate-invariant spectrum
$S_N(\omega)$ up to the aliasing term \eqref{eq:epsE}. The structure considered here
is restricted to integer $M$, but it keeps the same fixed, multiplier-free
hardware for all $M$.

Conventional droop compensators \cite{kim,dolecek,molnar,fernandez} act at
the low rate and are designed for a given $M$ or for the limit $M\to\infty$.
The derating filter is complementary to them: it acts at the high rate and
removes the reason why their coefficients would have to change with $M$.

\section{Conclusion}

The three-tap derating filter of \cite{lee} is the second-moment approximation
of an exact family, and coincides with it for $N=3$ and $N=4$. With the Eulerian filter $E_N$ a comb decimator
realizes the sampled B-spline of degree $N-1$ for every decimation factor.
The closed-form coefficients of maximally flat low-rate compensators then
depend on $N$ only, the group delay is $N/2$ low-rate samples for every $M$
and every $N$, and the residual rate dependency is the closed-form aliasing
sum \eqref{eq:epsE}, of order $M^{-N}$ or $M^{-(N+1)}$. The aim throughout is rate
invariance, not a flatter pass band: the droop $S_N(\omega)$ is left to the
compensator, which can now be designed once. For $N\le4$ the residual rate
dependency of the pass band is no smaller than with the three-tap design; for
$N\ge5$ it is lower by two or more orders in $M$. The exact filter also
extends rate independence from three-tap to longer compensators and removes
the bounds on $N$, at the cost of
$\lceil\log_2(N-1)!\rceil$ bits of word length. In short, a single fixed filter,
the Euler--Frobenius polynomial, makes both the low-rate compensator of a
multistage (IFIR-style) design and the polynomial (Farrow-type) description
of the comb simple and rate invariant. A comparison of hardware cost against sharpened and modified comb structures
is left for further work. Two further directions are treated separately: the
comb interpolator, where the exact and rate-independent values at the common
instants admit recursive realizations with reduced memory, and the
finite-word-length behaviour of such realizations.

\appendices
\section{Pass-Band Expansions}\label{app:a}

With $x=\omega/2M$, the deviation of the plain comb is
$\varepsilon^{\rm comb}=(x/\sin x)^{N}-1$. From
$\ln(x/\sin x)=x^{2}/6+x^{4}/180+O(x^{6})$,
\begin{equation*}
\Bigl(\frac{x}{\sin x}\Bigr)^{N}=1+\frac{N}{6}x^{2}
+\Bigl(\frac{N}{180}+\frac{N^{2}}{72}\Bigr)x^{4}+O(x^{6}),
\end{equation*}
which gives \eqref{eq:epscomb}. The three-tap filter \eqref{eq:DN} has the
zero-phase response $D_N=1-\tfrac{N}{12}(1-\cos2x)=1-\tfrac{N}{6}\sin^{2}x
=1-\tfrac N6x^{2}+\tfrac N{18}x^{4}+O(x^{6})$. Multiplying,
the $x^{2}$ terms cancel and the $x^{4}$ coefficient is
\begin{equation*}
\frac{N}{180}+\frac{N^{2}}{72}+\frac{N}{18}-\frac{N^{2}}{36}
=\frac{N(22-5N)}{360},
\end{equation*}
which is \eqref{eq:epsD} after substituting $x^{4}=(\omega/M)^{4}/16$. The
coefficient changes sign between $N=4$ and $N=5$; at $N=4$ it equals $1/720$,
in agreement with \eqref{eq:lead}, as it must since $D_4=E_4$.

\section{Compensator Coefficients}\label{app:b}

The compensator must equalize $S_N(\omega)$. With $x=\omega/2$ in the
expansion of Appendix~\ref{app:a},
\begin{equation*}
\frac{1}{S_N(\omega)}=1+\frac{N}{24}\,\omega^{2}
+\frac{N(5N+2)}{5760}\,\omega^{4}+O(\omega^{6}).
\end{equation*}
The symmetric five-tap filter with unit DC gain has the zero-phase response
\begin{align*}
C&=1-2c_a(1-\cos\omega)-2c_b(1-\cos2\omega)\\
&=1-(c_a+4c_b)\,\omega^{2}
+\Bigl(\frac{c_a}{12}+\frac{4c_b}{3}\Bigr)\omega^{4}+O(\omega^{6}).
\end{align*}
Equating the coefficients of $\omega^{2}$ and $\omega^{4}$ gives
\eqref{eq:c5}; with $c_b=0$ the $\omega^{2}$ equation alone gives
\eqref{eq:c0}. For the plain comb the target is
$\bigl(M\sin(\omega/2M)/\sin(\omega/2)\bigr)^{N}$, whose $\omega^{2}$
coefficient is $\tfrac N{24}(1-M^{-2})$; this gives \eqref{eq:c0M}, and the
$\omega^{4}$ coefficient likewise acquires terms in $M^{-2}$ and $M^{-4}$.

\end{document}